# Agptools: a utility suite for editing genome assemblies

Edward S. Ricemeyer[1,2], Rachel A. Carroll[1], Wesley C. Warren[1,3]

[1] Bond Life Sciences Center, University of Missouri, Columbia, MO, USA
[2] Palaeogenomics Group, Institute of Palaeoanatomy, Domestication Research and the History of Veterinary Medicine, Ludwig-Maximilians-Universität, Munich, Germany
[3] Division of Animal Sciences, Department of Surgery, University of Missouri, Columbia, MO, USA

**Background.** The AGP format is a tab-separated table format describing how components of a genome assembly fit together. A standard submission format for genome assemblies is a fasta file giving the sequence of contigs along with an AGP file showing how these components are assembled into larger pieces like scaffolds or chromosomes. For this reason, many scaffolding software pipelines output assemblies in this format. However, although many programs for assembling and scaffolding genomes read and write this format, there is currently no published software for making edits to AGP files when performing assembly curation.
**Results.** We present agptools, a suite of command-line programs that can perform common operations on AGP files, such as breaking and joining sequences, inverting pieces of assembly components, assembling contigs into larger sequences based on an AGP file, and transforming between coordinate systems of different assembly layouts. Additionally, agptools includes an API that writers of other software packages can use to read, write, and manipulate AGP files within their own programs.
**Conclusions.** Agptools gives bioinformaticians a simple, robust, and reproducible way to edit genome assemblies that avoids the shortfalls of other methods for editing AGP files.

**Introduction**

New technologies for genome sequencing and new algorithms for genome assembly have made it possible to assemble genomes to a chromosome level at a low cost (Rice and Green, 2019; Rhie *et al.*, 2021). The availability of highly accurate long reads (Cheng *et al.*, 2022), as well as ultra-long reads (Jain *et al.*, 2018), has even made gap-free telomere-to-telomere genome assemblies possible for an increasing number of species (Rautiainen *et al.*, 2023; Nurk *et al.*, 2022; Xue *et al.*, 2021; Bliznina *et al.*, 2021; Belser *et al.*, 2021; Olagunju *et al.*, 2024). Now, more and more species have many different genome assemblies available, which are then combined into pangenome graphs to provide advantages such as decreased reference bias, increased mapping accuracy, and better recovery of structural variation existing within the species (Groza *et al.*, 2024; Ebler *et al.*, 2022; Leonard *et al.*, 2023; Smith *et al.*, 2023).

However, although assembling high-quality genomes has become easier and more cost-effective over time, most assemblies today are a result of combining several types of data using a sequence of different software packages in a pipeline, and require human curation to correct errors between the steps. For example, a current best-practice genome assembly pipeline involves first assembling long reads into contigs, then combining the contigs into scaffolds using one or more long-range data sources such as optical mapping or proximity ligation, and finally assigning scaffolds into chromosomes based on either a physical map or synteny with another species, with manual curation occurring between steps to correct errors (Rhie *et al.*, 2021). The newer technique of telomere-to-telomere assembly uses programs that output an assembly graph rather than a linear assembly, which the user then has to curate (Rautiainen *et al.*, 2023; Cheng *et al.*, 2024).

Scaffolders such as SALSA (Ghurye *et al.*, 2019), YaHS (Zhou *et al.*, 2023), and RagTag (Alonge *et al.*, 2022) all output an AGP (A Golden Path) file that describes how the input sequences are combined and arranged into the larger output sequences. AGP is a tabular format where each line describes a segment of sequence in the output assembly, using coordinates of the input assembly (AGP Specification v2.1). For example, if an output scaffold is composed of two input contigs separated by a gap, the AGP file would use three lines to describe this scaffold: first, a line giving the coordinates and orientation of the first input contig in the output scaffold; next, a line describing the gap, including what evidence was used to infer the order and orientation of the sequences flanking the gap; and finally, a line describing the second input contig. These three lines, along with the sequences of the two contigs, thus contain all the information necessary to construct the resulting scaffolds.

The AGP file format has become a standard way of describing the layout of a genome assembly, and is the format for genome assembly submissions to both NCBI and Ensembl assembly databases. Representing an assembly as a set of contiguous sequences along with a description of how they are ordered and oriented into chromosomes has several advantages over representing it as a set of final sequences containing gaps represented as strings of N. An AGP contains more information, in the form of gap line evidence fields, about what data were used to join pairs of neighboring sequences, which is especially relevant when multiple data types were used to order and orient sequences. Perhaps more importantly, making changes to an assembly represented this way is much simpler. For example, to change the orientation of a contig in a fasta file of scaffolds, it is necessary to extract the sequence of this contig from its scaffold, reverse-complement the sequence, and then splice it back into the original file; however, performing the same operation in an AGP file is as simple as changing a "+" to a "-" on a single line.

Despite the wide use of AGP format among different genome assembly software, and the necessity to make edits to assemblies during assembly curation, there is not currently a published software package providing a simple way to make changes to an AGP file. Therefore, we developed agptools, a suite of command-line utilities that provides a simple interface for manipulating AGP files. This package also exposes an API that other developers can use to read, write, and manipulate these files in their own software, rather than needing to write their own code for these tasks.

**Availability**

Agptools is available under the open-source MIT License. It can be installed on any operating system running Python with one command. Most sub-commands can be run on a personal computer and require less than 1GB RAM. We commit to maintaining the code and documentation for two years after publication on GitHub at https://github.com/WarrenLab/agptools. It will be available permanently via Zenodo with DOI 10.5281/zenodo.15085069.

**Implementation**

Agptools is written in Python and can be run on Python versions >= 3.7. It uses the screed (screed: a simple read-only sequence database, designed for short reads) and pyfaidx (Shirley *et al.*, 2015) packages for fasta input and output. Its command-line interface consists of a single base command, "agptools", followed by a sub-command. Most subcommands take an AGP file and a second text file listing operations to perform as input, and output a modified AGP file. The subcommands are described below.

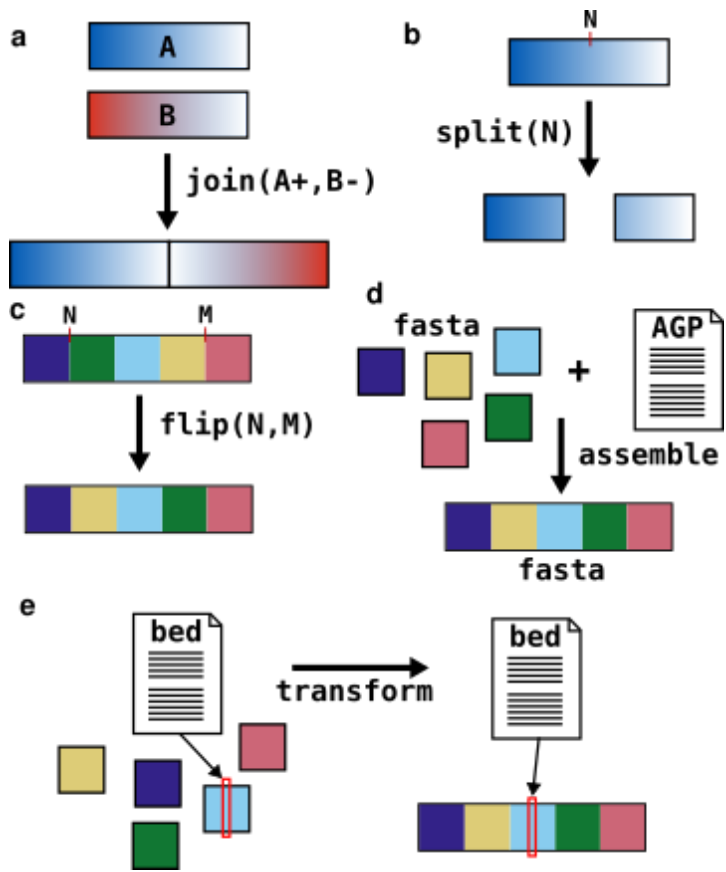

**Figure 1.** Subcommands of agptools. (a) The join command joins two or more separate sequences into a single sequence. (b) The split command splits a sequence at a given location into two sequences. (c) The flip command inverts part of a sequence. (d) The assemble command produces an assembled fasta file from a fasta file of sequence components and an AGP file giving their order and orientation. (e) The transform command translates coordinates in a bed file into the coordinate system of assembled sequences.

*Join*

The join operation takes two or more input sequences and joins them into a single output sequence, delimited by gaps (Figure 1a). It takes as input an AGP file and a list of joins to make, and outputs an AGP file with the requested sequences joined. The list of joins is a file where each line contains a list of input sequences to join into a single output sequence, along with their order and orientation. The size, type, and evidence listed on the newly created gap lines can be specified by optional arguments.

*Split*

The split operation takes a single sequence and breaks it into two or more sequences at specified breakpoints (Figure 1b). It takes as input an AGP file and a list of breakpoints, and outputs an AGP with the input sequences split at the given breakpoints. The list of breakpoints is a tab-separated file with two fields: the first is the name of the sequence to split and the second is a comma-separated list of breakpoints within that sequence. Sequences can be split both within gaps and within non-gap sequence.

*Flip*

The flip operation inverts part of a sequence (Figure 1c). It takes as input an AGP file and a list of segments to flip, and outputs an AGP file with the requested segments in the opposite orientation. The list of segments to invert is a tab-separated file with three columns: sequence name, start coordinate of segment to invert, and end coordinate of segment to invert.

*Remove*

The remove subcommand removes sequences from the input. This can be useful when curating an assembly to remove haplotigs or sequences representing contamination. It takes as input an AGP file and a list of names of sequences to remove, and outputs an AGP without the removed sequences.

*Rename*

The rename subcommand changes the names of input sequences. This can be useful for converting an assembly from generic computer-generated sequence names to chromosome or plasmid names. It takes as input an AGP file along with a tab-separated file with three columns: old sequence name, new sequence name, and new sequence orientation, and outputs an AGP file with sequences renamed.

*Assemble*

The assemble subcommand takes a fasta containing input sequences and an AGP file describing how these sequences are laid out into larger sequences, and outputs a fasta file with the input sequences laid out as described in the AGP (Figure 1d). This can be helpful for performing whole-genome alignment to use synteny with a related species for curation. Because whole contigs are loaded into memory in this command, it requires more RAM than the other subcommands.

*Transform*

Making changes to an assembly presents the problem of introducing new coordinate systems. For example, if reads were aligned to contigs, the coordinates of these alignments will not be the same once the contigs are assembled into scaffolds. The transform subcommand overcomes this problem by translating coordinates from one system to another (Figure 1e). As input, it takes a bed file with coordinates relative to the input sequences and an AGP that assembles input sequences into output sequences, and outputs a bed file with coordinates relative to the output sequences. The transform command is especially useful for translating bed coordinates of Hi-C alignments from contig coordinates to scaffold coordinates, so that heatmaps can be generated iteratively during curation without needing to realign Hi-C reads.

*Sanitize*

There are some constructs that are valid AGP format but not accepted by NCBI's assembly submission pipeline, such as single-component sequences with reverse-orientation and input contigs split among multiple output scaffolds. The sanitize subcommand fixes these issues so that an assembly can be submitted without running into problems.

*API*

All these operations, in addition to being usable via a command-line interface, can also be performed within external python code using the API. This API also exposes functions and data structures for input and output of files in AGP format. Therefore, the software can be used by other developers to robustly manipulate AGP files within their own programs.

*Documentation*

The documentation for this software package takes three forms. First, command-line help messages for the root command as well as each subcommand explain how to use the command-line interface, including input, output, and additional arguments where relevant. Second, a more comprehensive description of each command is hosted on github, with example input, command-line invocation, and output, as well as any additional notes on how it should be used. Finally, all classes and functions in the API are comprehensively documented in an API page.

*Testing*

Unit tests provide 100% code coverage of this software package using the pytest framework. All pull requests must pass all unit tests to be merged into the main branch of the repository.

**Discussion**

This software package provides a simple way for a user to manipulate genome assemblies during manual curation steps. It has already been used during the curation of several published genome assemblies (Warren *et al.*, 2024; Carroll *et al.*, 2024).

Agptools is written in Python, perhaps the most common programming language for bioinformatics, with MyPy type checking, linting, and autoformatting, and thorough source-code documentation. Moreover, numerous unit tests provide full code coverage of the package and decrease the likelihood that updates and bug fixes will introduce new bugs that break existing functionality. These factors, along with the simplicity of the software, make it highly maintainable.

One additional benefit of this software is that because all operations are performed as a single command with a file listing changes to make, assemblies curated using this package are highly reproducible. To reproduce the curation exactly, only the list of commands and the input files are necessary. This is different from curation software such as Juicebox Assembly Tools (Dudchenko *et al.*, 2018), where edits are made by clicking on a heatmap, and therefore, two people making the same series of edits are likely to produce slightly different results by clicking in slightly different places.

**Acknowledgements**

We thank all users of this software who have found and reported bugs.

**Author contributions**

ESR conceived the project with input from WCW. ESR wrote the software, documentation, and manuscript. RAC tested the software. All authors edited and approved the final manuscript.